\documentclass[conference]{IEEEtran}
\IEEEoverridecommandlockouts
\usepackage{cite}
\usepackage{amsmath,amssymb,amsfonts}
\usepackage{algorithmic}
\usepackage{graphicx}
\usepackage{textcomp}
\usepackage{xcolor}
\def\BibTeX{{\rm B\kern-.05em{\sc i\kern-.025em b}\kern-.08em T\kern-.1667em\lower.7ex\hbox{E}\kern-.125emX}}
\usepackage{array}
\usepackage[Symbol]{upgreek}
\usepackage{booktabs}
\usepackage{longtable}
\usepackage{mathtools}
\usepackage{tikz}
\usetikzlibrary{tikzmark}
\usetikzlibrary{arrows.meta}
\usepackage{multirow}
\usepackage{graphicx}
\usepackage{lscape}
\usepackage{float}
\usepackage{xcolor}
\usepackage[hidelinks]{hyperref}
\usepackage{dblfloatfix}
\usepackage{footnote}
\usepackage{makecell}
\usepackage{xcolor}
\usepackage[utf8]{inputenc}
\usepackage[automake,nonumberlist,nogroupskip,nopostdot]{glossaries}

\makeglossaries
\usepackage{forest,array}
\usepackage{tikz}
\usepackage{soul}
\usepackage{pifont}
\usepackage[parfill]{parskip}
\usepackage{titlesec}
\titlespacing*{\subsubsection}{0pt}{2ex plus 1ex minus .15ex}{.7ex plus .1ex}

\usepackage{diagbox}
\usepackage{pgfplots}
\pgfplotsset{width=7cm,compat=1.3}
\begin{document}
\setlength{\parskip}{0pt} 
\setlength{\parindent}{15pt}
\newacronym{vnf}{VNF}{Virtualized Network Function}
\newacronym{5g}{5G}{Fifth Generation}
\newacronym{mmtc}{mMTC}{massive Machine Type Communications}
\newacronym{embb}{eMBB}{enhanced Mobile Broadband}
\newacronym{urllc}{URLLC}{Ultra-Reliable Low Latency Communications}
\newacronym{sdn}{SDN}{Software Defined Networking}
\newacronym{nfv}{NFV}{Network Function Virtualization}
\newacronym{nf}{NF}{Network Function}

\newacronym{arpf}{ARPF}{Authentication credential Repository and Processing Function}

\newacronym{supi}{SUPI}{Subscription Permanent Identifier}
\newacronym{capex}{CAPEX}{Capital Expenditure}
\newacronym{opex}{OPEX}{Operational Expenditure}
\newacronym{dos}{DoS}{Denial-of-Service}
\newacronym{ddos}{DDoS}{Distributed DoS}
\newacronym{minlp}{MINLP}{Mixed-Integer Nonlinear Programming}
\newacronym{ilp}{ILP}{Integer Linear Programming}
\newacronym{upf}{UPF}{User Plane Function}
\newacronym{mec}{MEC}{Multi-access Edge Computing}
\newacronym{qos}{QoS}{Quality of Service}
\newacronym{sfc}{SFC}{Service Function Chain}
\newacronym{pdu}{PDU}{Protocol Data Unit}
\newacronym{ns}{NS}{Network Slicing}
\newacronym{sla}{SLA}{Service Level Agreement}
\newacronym{dps}{DPS}{Data Plane Services}
\newacronym{cps}{CPS}{Control Plane Services}
\newacronym{vm}{VM}{Virtual Machine}
\newacronym{mip}{MIP}{Mixed Integer Programming}
\newacronym{amf}{AMF}{Access and Mobility Function}
\newacronym{smf}{SMF}{Session Management Function}
\newacronym{nrf}{NRF}{Network Repository Function}
\newacronym{scip}{SCIP}{Solving Constraint Integer Programs}
\newacronym{ue}{UE}{User Equipment}

\newacronym{hplmn}{H-PLMN}{Home \gls{plmn}}
\newacronym{plmn}{PLMN}{Public Land Mobile Network}
\newacronym{eap-aka}{EAP-AKA'}{Extensible Authentication Protocol-Authentication and Key Agreement}
\newacronym{aka}{AKA}{Authentication and Key Agreement}
\newacronym{gnb}{gNB}{gNodeB}
\newacronym{3gpp}{3GPP}{3rd Generation Partnership Project}
\newacronym{ran}{RAN}{Radio Access Network}
\newacronym{udm}{UDM}{Unified Data Management}
\newacronym{ausf}{AUSF}{Authentication Server Function}

\title{ A Security-aware Network Function Sharing Model for 5G Slicing	\\
}
			
\author{
    \textbf{Authors’ draft for soliciting feedback, March 6, 2023}\\
    Mohammed Mahyoub, AbdulAziz AbdulGhaffar\textsuperscript{1}, Emmanuel Alalade, and	Ashraf Matrawy \\ School of Information Technology, Carleton University, Canada \\ \textsuperscript{1}Department of Systems and Computer Engineering, Carleton University, Canada 
				%
	}
			
\maketitle
\thispagestyle{plain}
\pagestyle{plain}
\begin{abstract}
	Sharing \glspl{vnf} among different slices in \gls{5g}  is a potential strategy to simplify the system implementation and utilize \gls{5g} resources efficiently. In this paper, we propose a security-aware \gls{vnf} sharing model for \gls{5g} networks. The proposed optimization model satisfies the service requirements of various slices, enhances slice security by isolating their critical \glspl{vnf}, and enhances resource utilization of the underlying physical infrastructure. The model tries to systematically decide on sharing a particular \gls{vnf} based on two groups of constraints;  the first group of constraints is common assignment constraints used in the existing literature. The second group is the novel security constraints that we propose in this work; the maximum traffic allowed to be processed by the \gls{vnf} and the exposure of the \gls{vnf} to procedures sourced via untrusted users or access networks. This sharing problem is formalized to allow for procedure-level modeling that satisfies the requirements of slice requests in \gls{5g} systems. The model is tested using standard \glspl{vnf} and procedures of the 5G system rather than generic ones. The numerical results of the model show the benefits and costs of applying the security constraints along with the network performance in terms of different metrics.
\end{abstract}
			
\begin{IEEEkeywords}
	\gls{5g} Security, \gls{ns}, \gls{vnf} Sharing, Optimization
\end{IEEEkeywords}

\section{Introduction}
    \IEEEPARstart{5G} networks are visioned to support various applications and services with diversified requirements \cite{stallings20215g}. One distinct concept in \gls{5g} architecture is the \acrfull{ns} which was not present in previous generations of cellular networks. \gls{ns} enables \gls{5g}  operators to deploy multiple logical networks on shared physical resources to serve traffic segments with different demands \cite{alliance2016description,foukas2017network}. This is achieved using different technologies integrated with \gls{5g} architecture such as, most notably,  \gls{nfv} technology. \gls{nfv} allows the deployment of \acrfullpl{vnf} in software or a virtualized environment on commodity hardware. Both \gls{ns} and \gls{nfv} help \gls{5g} operators to reduce the overall \gls{capex} and \gls{opex} by deploying \glspl{vnf} efficiently and flexibly to optimize the utilization of network resources  \cite{Samdanis2016}.
    		
    In this paper, we propose a security-aware \glspl{vnf} sharing model for \gls{5g}  networks. The proposed optimization model not only satisfies the service requirements of various slices but also enhances security by isolating their critical \glspl{vnf} while enhancing resource utilization of the underlying physical infrastructure. This goal is achieved by sharing as many noncritical \glspl{vnf} as possible to efficiently utilize resources and satisfy the latency limitations of the procedures composing  \gls{5g} slices. Although some literature studies considered the sharing property of \glspl{vnf} in the mapping process, they subjectively decide on this property and use it as an input to their model. This work tries to fill this gap by following a systematic way to decide whether a particular \gls{vnf} is critical and, if so, to avoid sharing it among slices.  In the proposed model, two novel security constraints are considered to define the \gls{vnf} criticality. The first constraint is the maximum traffic that can be processed by a particular \gls{vnf}. If a \gls{vnf} has to process large user and control traffic, it could become a bottleneck which makes it critical, and thus, should not be shared between slices.  The second one is exposure to procedures initiated by untrusted entities (i.e. user devices or networks). If a \gls{vnf} is exposed to procedures coming from untrusted parties, this \gls{vnf} should not be shared among slices too. In case of the exposed \gls{vnf} is compromised, this can impact all other slices that share it. To this end, providing isolation to critical \glspl{vnf} is very crucial in \gls{5g} network slicing. In light of the above discussion, the contributions of this paper are four-fold:
	\begin{itemize}
        \item Proposing a multi-objective \gls{minlp} model aiming at minimizing the processing capacity needed and procedures' latency of all requested slices.
		\item Providing a systematic way to decide on the sharing property of a particular \gls{vnf} by introducing new security constraints that define the \gls{vnf}'s criticality.
		\item Considering the procedure level granularity instead of abstracting a slice as a unit. To the best of our knowledge, this is the first work to consider procedure-level details in the optimization model.
        \item The proposed model is tested using standard  procedures and  \glspl{vnf} of \gls{5g} architecture that are discribed in \gls{3gpp} standards \cite{Specification2011} rather than using generic \glspl{vnf} or symbolic procedures.
	\end{itemize}

    The rest of this paper is structured as follows. Section \ref{sec:related} discusses the related works and Section \ref{sec:Model} explains the proposed model. The system setup and model parameters are presented in Section \ref{sec:sys_setup}. The standard \gls{5g} procedures implemented to test the model are defined in Section \ref{sec:5g_proc}. The proposed model is evaluated in Section \ref{sec:results} and the limitations of this work are provided in Section \ref{sec:limit}. Finally, Section \ref{sec:conclusion} concludes this study.
			
\section{Related work} \label{sec:related}
	This section reviews the related literature studies that attempted to solve the \gls{vnf} placement and allocation problem using optimization approaches. Although there are many studies that considered sharing a physical node between multiple \glspl{vnf}, our work mainly focuses on sharing the a \glspl{vnf} themselves between multiple \gls{5g} slices.

    Leyva \textit{et al.} in \cite{Leyva-Pupo2022} proposed an \gls{ilp}-formulated optimizing model for \glspl{upf}  chaining and placement in \gls{mec} system of \gls{5g}. Their model targeted the provisioning cost and \gls{qos} optimization. It considered several aspects such as resource capacity, service latency, \gls{upf}-specific requirements, and the order of \glspl{vnf} in the \glspl{sfc}.  \glspl{upf}  placement and routing are modelled as \gls{sfc} embedding problem in which active \gls{pdu} sessions are modelled as \gls{sfc} requests. There is no restriction on sharing a particular \gls{vnf} except its capacity limit. To solve the problem in a polynomial time, a customized heuristic along with simulating annealing algorithm has been proposed in their work. Our work, on the other hand, in addition to the data-plane function (i.e., \glspl{upf}), considers control-plane functions as well.
	
	Coelho \textit{et al.} in  \cite{Coelho2021, DaSilvaCoelho2020} modeled the provisioning of the \gls{ns} requests at the service level as an optimization problem. The model considered functional splitting in the radio access domain and also the separation of the control and data-plane functions. The authors assumed that the network slice request might impose constraints on  \glspl{vnf}  that can not be shared between \glspl{ns}  due to their criticality or their belonging to different tenants. They tested different sharing policies such as sharing \gls{dps} only, \gls{cps} only, some of \gls{dps}, some \gls{cps}, or without sharing constraints. These sharing policies are given to the model as input, however, our model decides whether to share \gls{vnf} systematically based on different security constraints. 
	
	Malandrino \textit{et al.} \cite{Malandrino2019b} studied reducing the cost of the \gls{5g} service deployment through sharing \glspl{vnf} subject to end-to-end delay requirements. With the assumption that there is no isolation needed for the new service request, \glspl{vnf} are shared if convenient (i.e. meet the delay requirements). The authors focused more on how to assign priorities for traffic flows that share the same \gls{vnf}. For this, they randomly assigned flows priority upon entering \glspl{vnf}. To reduce the time complexity, they proposed FlexShare as an assignment algorithm. 
	
	Tang \textit{et al.} \cite{Tang2019} proposed a dynamic scaling approach for \gls{vnf} based on traffic analysis and \gls{vnf} placement. They analyzed the traffic characteristic of operator networks and then proposed an organizational approach for \gls{vnf} placement in a common data center. Their model aims to achieve high service availability and save computational resources depending on the traffic estimation to scale in/out the \gls{vnf} instance. The authors considered general \gls{vnf} and only user traffic. Our work, however, considers the actual \glspl{vnf} of the \gls{5g} core, data plane, and control plane traffic.
	
	Truong-Huu \textit{et al.} in \cite{Truong-Huu2019b} leveraged the \gls{vnf}'s sharing property in their optimization model to minimize the bandwidth and computational resources required to serve slice requests. A \gls{vnf} is identified as shareable depending on its functionalists so that it can be assigned to serve different slices. The network address translation function is an example of shareable \gls{vnf}, however, firewall service is non-shareable. In their work, the sharing property of a \gls{vnf} is set in advance and provided to the model as input. Additionally, their work considered random traffic flows and generic \glspl{vnf} that serve these traffics.  
	
	Another work leveraging the shareable \glspl{vnf} criteria to enhance resource utilization is presented by Chengli  \textit{et al.} in \cite{Mei2020b}. Their enhancement is evaluated in terms of the slice acceptance ratio. They claimed that some common functions such as mobility management and network address translation functions  can be shared across multiple slices. Similar to the work in  \cite{Truong-Huu2019b}, Chengli \textit{et al.} \cite{Mei2020b} randomly set the sharing property in their experiments and used it as an input for their model.
	
	A queuing-based system model is proposed by Agarwal \textit{ et al.} \cite{Agarwal2018a}  for optimizing the \glspl{vnf} placement and allocation in physical hosts taking into account the \glspl{vnf} sharing. Authors in \cite{Agarwal2018a} utilized the concept of queuing theory and considered random procedures with a sequence of random and generic \glspl{vnf}. In our paper, we consider \gls{5g} \glspl{vnf} and multiple standard \gls{5g} procedures.
	Other models were presented in \cite{Golkarifard2021, Luizelli2017} to optimize the utilization of the underlined physical infrastructure considering different slicing requirements. However, both of them assumed that \glspl{vnf} can't be shared among slice or service requests. 
 
 Finally, Sattar  \textit{et al.} proposed an optimal slice allocation model in 5G core networks \cite{Sattar2019} and extended it to propose a security-aware optimization model to protect the \gls{5g} core network slices against \gls{ddos} attacks in \cite{danish2019}. The model tried to isolate the network slices at the hardware level. The authors considered both inter-slice and intra-slice isolation and evaluated the performance of their proposed solution on a testbed which involved both simulation and experimental parts. Their results confirmed the benefits of utilizing a security-aware network slice optimization model to mitigate the impact of \gls{ddos} attacks. Our work focuses on sharing and isolation of \gls{5g} \glspl{vnf} whereas the work presented in \cite{danish2019} only considers sharing of physical resources and not the \glspl{vnf}. Furthermore, our work considers the standard \glspl{vnf} of the \gls{5g} core along with several procedures used in the \gls{5g} network.
	
	To sum up, it can be observed that the capacity limit of the \glspl{vnf} is the thing that we have in common with most of the literature studied which is standard in this area. To the best of our knowledge, this is the first study incorporating security aspects into the optimization model. Not only that, but our work also proposes a systematic way to decide on whether to share or not to share a particular \gls{vnf} installed in a specific physical node. Additionally, our work considers procedure-level rather than slice requests or traffic flows, and some standard \glspl{vnf} and procedures of \gls{5g} rather than generic or symbolic ones.

	\section{System Model and Problem Formulation} \label{sec:Model}
	In this section, the proposed model is given.  The \gls{vnf} sharing problem considered in this study is formalized and solved as an \gls{minlp}. The proposed model aims to optimize computational processing costs and the latency of slices' procedures. 
	
	\subsection{Model Description and Notations}
	The modeling of the virtual and physical networks is defined in this subsection. In this model, each slice request $s \in \mathcal{S}$  is composed of a set of procedures, $\mathcal{P}_s$. The virtual network is modeled in this work by a set of directed graphs. Each graph $(\mathcal{V}^s_p, \mathcal{R}_p^s)$ corresponds to a particular procedure $p \in \mathcal{P}_s $ that belongs to a specific slice $s \in \mathcal{S}$, where $\mathcal{V}^s_p$ is the set of \glspl{vnf} serving the procedure $p$ and  $\mathcal{R}_p^s$ denotes the set of the virtual links used by that procedure.  Each procedure $p \in \mathcal{P}_s$ requires a specific data rate, $\lambda_p^s~$, and a maximum tolerated delay,   $\delta_p^{s,max}$. Each \gls{vnf} $v \in \mathcal{V}$ is represented by a tuple  $ \langle v_{i},~I_{v},~\delta^n_{v_i},~\zeta^{max}_{v},~\mu_{v},~\omega_{v} \rangle $  where $v_{i}$ is the deployed  $i^{th}$  instance of \gls{vnf} $v$ type, $I_{v}$ denotes the set of all instances of \gls{vnf} $v$ type deployed across all physical nodes, $\delta^n_{v_i}$ is the processing delay for $i^{th}$ instance of type $v$ deployed in node $n$, $\zeta^{max}_{v}$ is the maximum accepted processing capacity to which the \gls{vnf} type $v_i$ can be extended, $\mu_{v}$ is the per unit processing capacity required by the \gls{vnf} of type $v$, and $\omega_{v}$ denotes the number of processed data units per unit processing time by the \gls{vnf} of type $v$. Finally, the physical infrastructure network is modeled as a directed graph $\mathcal{G}=(\mathcal{N}, \mathcal{L})$, where $\mathcal{N}$ is the set of physical nodes and   $\mathcal{L}$ denotes the physical links between these nodes. Each physical node $n \in \mathcal{N}$ has a finite processing capacity, $C^{max}_{n}$. A physical link $(n,m)$, between node $n$ and node $m$,  entails a deterministic delay $d{(n,m)}$ proportional to its length and also a maximum bandwidth capacity $b{(n,m)}$.   Table \ref{tab:analyticalSymbols} summarizes the notations and variable definitions used throughout this paper.
 \begin{table}[h!]
	\centering
	\caption{Used notations summary}
	\label{tab:analyticalSymbols}
	\begin{tabular}{m{1.3cm} m{7cm} }	\hline 
		\textbf{Parameter} & \textbf{The definition}\\
		\hline  \hline 
		$\mathcal{S}$& The set of all slices \\
		$\mathcal{P}$& The set of all procedures \\
		$\mathcal{V}$& The set of all \gls{vnf} types \\
		$\mathcal{N}$& The set of all physical nodes\\ 
		$\mathcal{L}$& The set of all physical connections between nodes in $\mathcal{N}$ \\ 
		$(n,m)$& The physical link beginning at node $n$ and ending at node  $m$\\
		$d{(n,m)}$ & The delay of the physical link $(n,m)$ \\
		$b{(n,m)}$&The bandwidth capacity of the physical link $(n,m)$ \\
		$\mathcal{R}$&The set of virtual links used by all procedures in the network\\ 
		$\mathcal{P}_s$& The set of all procedures belong to the slice $s$ \\
		$\mathcal{V}^s_p $& The set of all \glspl{vnf} belong to procedure $p$ in slice $s$ \\
		$v_{i}$&The deployed  $i^{th}$  instance of \gls{vnf} $v$ type	\\
		$I_{v}$&The set of all instances of \gls{vnf} $v$ type deployed over all physical nodes	\\
		$(v_i,z_j)$& The virtual link between \gls{vnf} instances $v_i$ and $z_j$  \\
		$\mathcal{R}_p^s$&The set of virtual links used by procedure $p$ of slice $s$\\ 
		$\lambda_p^s$&The packet rate of the procedure $p$  given one \gls{ue} is connected to the $s$\\
		$\delta^n_{v_i}$&The processing delay   for $i^{th}$ instance of the type $v$ deployed in node $n$	\\
		$\delta_p^{s,max}$ & The maximum tolerated delay of the procedure $p$ \\
		$C^{max}_{n}$&The maximum processing capacity  of the node $n$		\\
		$\zeta^{max}_{v}$&The maximum accepted processing capacity to which the \gls{vnf} type $v_i$ can be extended		\\
		$\mu_{v}$&Per unit processing capacity required by the \gls{vnf} $v$ type	\\
		$\omega_{v}$&Number of processed data units per unit processing time by the \gls{vnf} $v$ type	\\
		$\delta_p^s$ & The delay of the procedure $p$ of slice $s$ based on the existing configuration \\
		$\zeta^n_{v_i}$&The total required processing capacity   of \gls{vnf} instance $v_i$ deployed in node $n$	\\
		$\psi_p^s$ & Indicates whether the procedure $p$ is sourced by external entity  \\
		$\eta^{s}_{v,p}$ &  indicates whether the \gls{vnf}-type $v$ is the first \gls{vnf} traversed by procedure $p$ \\
		$\theta^s_{v,p}$ &Indicates whether a procedure must traverse a \gls{vnf} type $v$ \\\\	\hline 
		
		\textbf{Variable}& \textbf{The definition}\\		\hline 
		$\upgamma^{n,s}_{v_i,p}$ & Binary variable indicates whether \gls{vnf} $v_i$ used by the procedure $p$  of slice $s$ is deployed at the node $n$ \\
		$\chi^{(n,m),s}_{(v_i,z_j),p}$ & Binary variable indicates whether the virtual link $(v_i,z_j)$, used by procedure $p$ of slice $s$, is mapped to the physical link $(n,m)$ \\
		$\beta^{n}_{v_i}$ & Binary variable indicates whether the \gls{vnf} instance $v_i$ is deployed at the node $n$ \\
		$\Omega^{n,s}_{v_i}$ & Binary variable that indicates whether the \gls{vnf} instance $v_i$ deployed in physical node $n$ is exposed to the outside by the slice $s$ \\

		\hline
	\end{tabular}
    \end{table}
    
\subsection{Model Assumptions}
    Few assumptions are considered in this work as outlined in this subsection. The standard \gls{5g} \glspl{vnf} considered in the model such as \gls{amf}, \gls{smf}, \gls{nrf}, etc.  could be VM-based or container-based  \glspl{vnf}. It is not important that the number of \glspl{vnf} type per slice is the same as that for another slice. Additionally, multiple instances of the \gls{vnf} type can be initiated if required as assumed in \cite{Golkarifard2021}. \glspl{vnf} are required to dynamically support scale-in and scale-out with minimal impact on the service quality offered \cite{ITU-T2018}. 
    Physical nodes are geographically distributed and each of them can deploy any \gls{vnf} type. 
    It is assumed that all traffic units need the same computational capability for processing. Although in this work we focus on CPU or computational capacity, the storage and memory could be accommodated.

     In this model, the delay raised by the load balance in the case of muli-core \gls{vnf} is minimal to be considered. The load balancer is needed when a \gls{vnf} requires processing capabilities that cannot be fulfilled by a single core. In this case, multiple cores are needed to satisfy the processing requirement of that \gls{vnf}. The load balancer will be used to balance the traffic between the cores and may lead to some performance penalties.
     Additionally, the context switching delay  caused by sharing the CPU's cores  between multiple \glspl{vnf} is not taken into account. This delay comes in a form of cache sharing and saving/loading the context of different \glspl{vnf}. It is linearly increased with the number of procedures using those \glspl{vnf}.

\subsection{The Objective Function}
	The first part of the objective function of this model is to minimize the total processing capacity needed to serve all slices. This part is satisfied by sharing as many noncritical \glspl{vnf} as possible while considering the security constraints imposed to mitigate the risks that raise by such sharing.  The second part is to minimize the delay of all procedures. These two parts can be formulated in Eq.(\ref{eq:obj-func}). It is worth mentioning here that we use common optimization goals in the literature which are minimizing delay and resource consumption.
    Although we focus on minimizing the processing capacity and procedures delay, our model can be extended to consider additional key performance indicators seamlessly.
			
	\begin{equation}
	\underset{\upgamma^{n,s}_{v_i,p}}{\min}  \sum_{v \in \mathcal{V}} \sum_{i \in I_v} \sum_{n \in \mathcal{N}} 	\tikzmarknode{left}  {\zeta^n_{v_i}} + \sum_{s \in \mathcal{S}}  \sum_{p \in \mathcal{P}_s} \tikzmarknode{right} {\delta^s_p}
	\begin{tikzpicture}[overlay,remember picture,red,>=stealth,shorten
		<=0.2ex,nodes={font=\small,align=left,inner ysep=1pt},<-]
		\path (left.south) ++ (0,-1.2em) node[anchor=north east] (diff)
		{The total required computational \\ capacity for all \glspl{vnf}};
		\draw (left.south) |- ([xshift=0.3ex]diff.south west);
		\path (right.south) ++ (0,-.2em) node[anchor=north west] (diff)
		{The total delay\\ of all procedures};
		\draw (right.south) |- ([xshift=0.3ex]diff.south east);
	\end{tikzpicture}
    \label{eq:obj-func}
    \end{equation}
    $~$\\ \\ \\
	\textbf{Subject to} constraints (\ref{eq:domain-constraints}) to (\ref{eq:exposure-constrnt}). 
 
\subsubsection{Computational Capacity and Procedure Delay}
In the following, we show how the computational capacity needed for a particular \gls{vnf} and  a procedure delay are calculated.
\begin{itemize}
    \item \textit{\gls{vnf} computational capacity}:
    Generally, the more services/procedures a \gls{vnf} provides/hosts, the more physical resources are required. The processing capacity $\zeta^n_{v_i}$,  that is needed by a particular  \gls{vnf}, comes in two forms; operational or base processing capacity $\zeta^{n,B}_{v_i}$ and traffic processing capacity $\zeta^{n,T}_{v_i}$ as shown in Eq. (\ref{eq:vnf-total-capacity}). Based on the number of procedures that a particular \gls{vnf} $v$ instance serves, we can calculate its $\zeta^{n,T}_{v_i}$. If the \gls{vnf} type $v$ requires $\mu_v$ processing capability to process one unit of traffic, then the $\zeta^{n,T}_{v_i}$ calculated as in Eq. (\ref{eq:vnf-traffic-capacity}). 
	\vspace{.5cm}
	\begin{equation}
    	\zeta^n_{v_i} = \tikzmarknode{left} {\zeta^{n,B}_{v_i}} .~ \tikzmarknode{mid}{\beta^{n}_{v_i}} +  \tikzmarknode{right} {\zeta^{n,T}_{v_i}} ~~~~~\forall v \in \mathcal{V}, ~~ i \in I_v,~  n \in \mathcal{N}
    	\begin{tikzpicture}[overlay,remember picture,red,>=stealth,shorten
    		<=0.2ex,nodes={font=\small,align=left,inner ysep=1pt},<-]
    		\path (left.north) ++ (0,-1.5em) node[anchor=north east] (diff)
    		{The $v$'s base capacity};
    		\draw (left.south) |- ([xshift=0.5ex]diff.south west);
    		\path (mid.south) ++ (0,-.2em) node[anchor=north west] (diff)
    		{Is the \gls{vnf} instance $v_i$ activated?};
    		\draw (mid.south) |- ([xshift=0.3ex]diff.south east);
    		\path (right.north) ++ (0,2em) node[anchor=north west] (diff)
    		{The traffic's processing  capacity};
    		\draw (right.north) |- ([xshift=0.3ex]diff.south east);
    	\end{tikzpicture}
    	\label{eq:vnf-total-capacity}
    \end{equation}
    $~$\\ 
	The total processing capacity than can be calculated as in Eq. (\ref{eq:vnf-total-capacity})
	$~$\\ 
    \begin{equation}
		\zeta^{n,T}_{v_i}~=~
		\sum_{s \in \mathcal{S}}  \sum_{p \in \mathcal{P}_s}~ 
		\tikzmarknode{left}{~\lambda_p^s} 
		\tikzmarknode{mid}{~\upgamma^{n,s}_{v_i,p} } 
		\tikzmarknode{right}{~\mu_{v}} \\
		~~~~~\forall v \in \mathcal{V}, ~~ i \in I_v,~  n \in \mathcal{N}
		\begin{tikzpicture}[overlay,remember picture,red,>=stealth,shorten
			<=0.2ex,nodes={font=\small,align=left,inner ysep=1pt},<-]
			\path (left.north) ++ (0,2em) node[anchor=north west] (diff)
			{The procedure's  packet rate};
			\draw (left.north) |- ([xshift=0.3ex]diff.south east);
			\path (right.south) ++ (0,-1.2em) node[anchor=north east] (diff)
			{The capacity needed for one traffic unit};
			\draw (right.south) |- ([xshift=0.3ex]diff.south west);
		\end{tikzpicture}
		\label{eq:vnf-traffic-capacity}
	\end{equation}
	$~$ \\ \\
    
    \item \textit{Procedure delay}:
    The experienced delay by a particular procedure $p$ is calculated from two parts. The first part is the processing delay incurred by \glspl{vnf} that the procedure passes through. The second part is the propagation delay of the links that the procedure uses. The total delay can be calculated as in Eq. (\ref{eq:procedure-delay})  
    
    \begin{equation}
    \begin{split}
        \delta^s_p =  \sum_{v \in \mathcal{V}^s_p} \sum_{i \in I_v} \sum_{n \in \mathcal{N}} \tikzmarknode{left} {\delta^n_{v_i}} ~ \upgamma^{n,s}_{v_i,p}  \\  \\~+~  \sum_{ (v_i, z_j) \in \mathcal{R}_p^s} \sum_{(n,m) \in \mathcal{L}}  \tikzmarknode{right} {d(n,m)} ~ \chi^{(n,m),s}_{(v_i,z_j),p}
    	~~~~~\forall s \in \mathcal{S},   p \in \mathcal{P}_s
    \begin{tikzpicture}[overlay,remember picture,red,>=stealth,shorten
    	<=0.2ex,nodes={font=\small,align=left,inner ysep=1pt},<-]
    	\path (left.south) ++ (0,-1.2em) node[anchor=north east] (diff)
    	{The delay of the instance of type $v$};
    	\draw (left.south) |- ([xshift=0.3ex]diff.south west);
    	\path (right.south) ++ (0,-1em) node[anchor=north west] (diff){The delay of the link (n,m)};
    	\draw (right.south) |- ([xshift=0.3ex]diff.south east);
    \end{tikzpicture}
    	\end{split}
    \label{eq:procedure-delay}
    \end{equation}
    Where the delay that is incured by a particular \gls{vnf} instance $v_i$ is calculated by Eq.(\ref{eq:vnf-processing-delay}).
    \begin{equation}
    \begin{split}
    	\delta^n_{v_i} =  1/\omega_{v} ~+~   1/(\omega_{v} - \sum_{s \in \mathcal{S}}  \sum_{p \in \mathcal{P}_s} \lambda_p^s ~ \upgamma^{n,s}_{v_i,p}) \\
    	~~~~~\forall s \in \mathcal{S},   p \in \mathcal{P}_s
    \end{split}
    \label{eq:vnf-processing-delay}
    \end{equation}
\end{itemize}  
    
    
    The remaining part of this section explains the model constraints. They are categorized under two groups;  assignment, and security constraints as explained in the following two subsections.
\subsubsection{Assignment constraints}
    We use common assignment constraints existing in many literature papers such as \cite{Leyva-Pupo2022} and \cite{Mei2020b}. 
    We develop assignment constraints along the same lines as other references, however, we develop these constraints to consider procedure level, specific \glspl{vnf} type, and both control- and data-plane functions. In the following points, the assignment constraints are given.
    \begin{itemize}
        \item Firstly, Eq. (\ref{eq:domain-constraints}) defines the value constraint of variables used in this model.
        \begin{equation}
        	\chi^{(n,m),s}_{(v_i,z_j),p}~, ~ \beta^{n}_{v_i}~, ~ \upgamma^{n,s}_{v_i,p}~, ~ \psi_p^s, ~\Omega^{n,s}_{v_i} ~ \in \{0,1\} 
        	\label{eq:domain-constraints}
        \end{equation}
        \item Let's denote by $\mathcal{P}_s$ the set of \glspl{vnf} required by the procedure $p$ of the slice $s$.  Constraint (\ref{eq:all-vnf-deployed}) guarantees that each procedure and its respective \glspl{vnf} are mapped.  
        \begin{equation}
        	\sum_{  n \in \mathcal{N}} \sum_{i \in I_v}   \upgamma^{n,s}_{v_i,p} = \tikzmarknode{right} {\theta^s_{v,p}} ~~~~~\forall s \in \mathcal{S}, ~ p \in \mathcal{P}_s, v \in \mathcal{V}
        	\label{eq:all-vnf-deployed}
        	\begin{tikzpicture}[overlay,remember picture,red,>=stealth,shorten
            	<=0.2ex,nodes={font=\small,align=left,inner ysep=1pt},<-]
            		\path (right.south) ++ (0,-.6em) node[anchor=north west] (diff){Must the procedure $p$ traverse the \gls{vnf} $v$?};
            		\draw (right.south) |- ([xshift=0.3ex]diff.south east);
            \end{tikzpicture}
        \end{equation}
    $~$\\
				
    				

    	\item Constraint (\ref{eq:vnf-launch-constrnt1}) and (\ref{eq:vnf-launch-constrnt2}) ensure that a \gls{vnf} instance will not be initiated unless there is at least one procedure using it. 
		    \begin{equation}
				\sum_{  s \in \mathcal{S}} \sum_{p \in \mathcal{P}_s}   \upgamma^{n,s}_{v_i,p} \le \mathcal{M} ~ \beta^{n}_{v_i} , ~~~~~\forall v \in \mathcal{V}, ~ i \in I_v, ~ n \in \mathcal{N}
				\label{eq:vnf-launch-constrnt1}
			\end{equation}
			Where $\mathcal{M}$ is a parameter greater than the maximum number of procedures that will be mapped to the instance $v_i$.
			\begin{equation}
    			 \beta^{n}_{v_i} - 	\sum_{  s \in \mathcal{S}} \sum_{p \in \mathcal{P}_s}   \upgamma^{n,s}_{v_i,p} <= 0, ~~~~~\forall v \in \mathcal{V}, ~ i \in I_v, ~ n \in \mathcal{N}
    			\label{eq:vnf-launch-constrnt2}
		    \end{equation}
				
		%
				
		\item Constraint (\ref{eq:vnf-instance-constrnt}) ensures that each instance $i \in I_v$ of a \gls{vnf} type $v$ is installed on one physical node at most.
		\begin{equation}
			\sum_{i \in I_v} \beta^{n}_{v_i} \le 1, ~~~~~\forall v \in \mathcal{V}, ~n \in \mathcal{N} 
			\label{eq:vnf-instance-constrnt}
		\end{equation}
				
%
	    \item The total capacity of a particular \gls{vnf} instance,  needed to process all procedures mapped to it, cannot exceed the absolute maximum capacity assigned to that \gls{vnf}. This constraint has been considered in other papers in the literature such as \cite{Malandrino2019b} and \cite{Leyva-Pupo2022}.
		\begin{equation}
			\zeta^n_{v_i} \le \tikzmarknode{right} {\zeta^{max}_{v}}, ~~~~~\forall n \in \mathcal{N},~~  v \in \mathcal{V}, ~~ i \in I_v
			\begin{tikzpicture}[overlay,remember picture,red,>=stealth,shorten
    			<=0.2ex,nodes={font=\small,align=left,inner ysep=1pt},<-]
    				\path (right.south) ++ (0,-1em) node[anchor=north west] (diff){The maximum computational capacity \\ assigned to the \gls{vnf} $v$};
    				\draw (right.south) |- ([xshift=0.3ex]diff.south east);
		    \end{tikzpicture}
			\label{eq:vnf-capcty-limt}
		\end{equation}
		$~$\\ \\
				
		\item Constraint (\ref{eq:node-capcty-limt}) ensures that the total capacity used by all \glspl{vnf} deployed in a physical node $n$ does not exceed the maximum processing capacity of that node.
		\begin{equation}
    		\sum_{v \in \mathcal{V}} \sum_{i \in I_v} \zeta^n_{v_i} ~.~	\beta^{n}_{v_i}  ~~\le C^{max}_{n}, ~~~~~\forall n \in \mathcal{N} 
    		\label{eq:node-capcty-limt}
		\end{equation}
	\end{itemize}
			
    \begin{itemize}
		\item Constraint (\ref{eq:phy-to-vir-links-mapping}) ensures that a physical link $(n,m)$ is used by a particular procedure, to map virtual link $(v_i, z_j)$, iff the two \glspl{vnf} $v_i$ and $z_j$ are mapped to nodes $n$ and $m$, respectively. This constraint is a non-linear constraint.
		\begin{equation}
			\begin{split}
				\chi^{(n,m),s}_{(v_i,z_j),p} \le \upgamma^{n,s}_{v_i,p} ~ .~ \upgamma^{m}_{z_j,p}  ~ \\ 
				\forall   (m, n) \in \mathcal{L}, ~ (v_i, z_j) \in \mathcal{R}, ~ p \in \mathcal{P}
			\end{split}
			\label{eq:phy-to-vir-links-mapping}
		\end{equation}

    	\item Constraint (\ref{eq:link-capcity-constrnt}) ensures that the total bandwidth required by all procedures that move between \glspl{vnf} through a particular link, $(n,m)$, are limited by the finite capacity of that link, $\zeta^{max}_{(n,m)}$ 
    	\begin{equation}
    		\begin{split}
    			\sum_{s \in \mathcal{S}}  \sum_{p \in \mathcal{P}_s} \sum_{ (v_i, z_j) \in \mathcal{R}_p^s}  \lambda^s_p ~~ \chi^{(n,m),s}_{(v_i,z_j),p} ~\le ~ b(n,m) \\ ~~ \forall (n,m) \in \mathcal{L}
    		\end{split}
    		\label{eq:link-capcity-constrnt}
    	\end{equation}
				
	    \item Constraint (\ref{eq:procedure-delay-constrnt}) certifies that the latency introduced by nodes' processing and network propagation can't exceed the maximum tolerated latency of a particular procedure.
	\end{itemize}
    \begin{equation}
		\delta^s_p \le \delta^{s,max}_p ~ ~~~~~\forall   s \in \mathcal{S}, ~ p \in \mathcal{P}_s
		\label{eq:procedure-delay-constrnt}
	\end{equation}
			
\subsubsection{Security constraints}
	Two security constraints are formulated in the model; \gls{vnf}'s maximum traffic and \gls{vnf} exposure constraints. These constraints are explained as follows:
	\begin{itemize}
	    \item \textit{The \gls{vnf}'s maximum traffic:} 
        This constraint ensures that the traffic processing capacity $\zeta^{T, n}_{v_i}$ of a \gls{vnf} instance $v_i$ should not exceed the predefined maximum traffic processing capacity  $\zeta^{T, max}_{v}$. Using this constraint, the $\zeta^{T, max}_{v}$ for a critical \gls{vnf} instance can be set at a lower value, and hence it will not be shared which will protect the critical \gls{vnf}. This is represented in constraint  (\ref{eq:vnf-traffic-limt1}).
			
		\begin{equation}
    		\zeta^{n,T}_{v_i} ~  \leq \zeta^{T, max}_{v}, ~~~~~\forall v \in \mathcal{V}, ~~ i \in I_v, ~  n \in \mathcal{N}
    		\label{eq:vnf-traffic-limt1}
	    \end{equation}
				
		\item  \textit{The \gls{vnf} exposure:} 
		The \gls{vnf} that is exposed to the outside network cannot be assigned to more than one slice. A \gls{vnf} is exposed to the outside network if it the first \gls{vnf} in the \glspl{vnf} chain serving a procedure that is initiated by the \gls{ue} or the \gls{ran}.  First, let $\Omega^{n,s}_{v_i}$ denotes to that \gls{vnf} instance $v_i$ deployed in physical node $n$ is exposed to the outside by the slice $s$. The $\Omega^{n,s}_{v_i}$ is calculated by Equations~(\ref{eq:exposure-calculation1}) and (\ref{eq:exposure-calculation2}) 
		$~$\\
        				
		\begin{equation}
    		\begin{split}
        	\sum_{p \in \mathcal{P}_s}     \tikzmarknode{left} {\eta^{s}_{p,v}} ~ \tikzmarknode{mid} {\psi^s_p} ~ \upgamma^{n,s}_{v_i,p} \le \mathcal{C} ~ \Omega^{n,s}_{v_i}  \\ \forall s \in \mathcal{S}, v \in \mathcal{V}, ~ i \in I_v, ~ n \in \mathcal{N}
        	\begin{tikzpicture}[overlay,remember picture,red,>=stealth,shorten
        	<=0.2ex,nodes={font=\small,align=left,inner ysep=1pt},<-]
        	    \path (left.south) ++ (0,-2.5em) node[anchor=north west] (diff){Is the \gls{vnf} $v$ the first on the \\ \glspl{vnf} sequence of the procedure?};
        		\draw (left.south) |- ([xshift=0.3ex]diff.south east);
        		\path (mid.north) ++ (0,3em) node[anchor=north west] (diff)
        		{Is the procedure $p$ sourced \\ externally?};
        		\draw (mid.north) |- ([xshift=0.3ex]diff.south east);
            \end{tikzpicture}
    	    \end{split}
            \label{eq:exposure-calculation1}
        \end{equation}
		$~$\\ \\ \\
		Where $\mathcal{C}$ is a parameter greater than the maximum number of procedures mapped into the $v_i$ and sourced externally.
		\begin{equation}
			\begin{split}
				\Omega^{n,s}_{v_i} - \sum_{p \in \mathcal{P}_s}     \eta^{s}_{p,v} ~\psi^s_p ~ \upgamma^{n,s}_{v_i,p} \le 0   \\ \forall s \in \mathcal{S}, v \in \mathcal{V}, ~ i \in I_v, ~ n \in \mathcal{N}
			\end{split}
			\label{eq:exposure-calculation2}
		\end{equation}
				
		Then the constraint (\ref{eq:exposure-constrnt}) ensures that the \gls{vnf} instance $v_i$ must not assigned to more than one slice.
				
		\begin{equation}
			\sum_{s \in \mathcal{S}}   \tikzmarknode{left} {\Omega^{n,s}_{v_i}} \le 1   ~~~~~\forall v \in \mathcal{V}, ~ i \in I_v,~ n \in \mathcal{N}
			\begin{tikzpicture}[overlay,remember picture,red,>=stealth,shorten
				<=0.2ex,nodes={font=\small,align=left,inner ysep=1pt},<-]
				    \path (left.south) ++ (0,-0.5em) node[anchor=north west] (diff){Indicating whether $v_i$ is exposed externally};
					\draw (left.south) |- ([xshift=0.3ex]diff.south east);
			\end{tikzpicture}
			\label{eq:exposure-constrnt}
		\end{equation}
	\end{itemize}

\section{System Setup}\label{sec:sys_setup}
    This section provides details about the system setup and values of the parameter used to test the model. 
    
    \textbf{The used solver:} The proposed model is implemented using the JuMP modeling language \cite{DunningHuchetteLubin2017} which is embedded in Julia \cite{bezanson2017julia}. As our model contains a combination of linear as well as non-linear constraints,  the \gls{scip} solver \cite{scip, scipopt} is employed  to solve the modeled problem. The \gls{scip} is currently one of the fastest non-commercial solvers available to solve problems of  \gls{mip} and \gls{minlp}  classes \cite{scip}. The experiments are performed on a Linux machine that has an Intel processor with $32$ cores and $32$ GB of RAM.
    
    \textbf{The environment set-up:} A total of three simulated physical nodes with a maximum of $30$ capacity units each are considered in these experiments. Each \gls{vnf} requires one capacity unit of the physical node to be deployed (or activated) and one more capacity unit to serve one procedure in each traverse. For instance, if a procedure needs to use a \gls{vnf} more than once, then the \gls{vnf} will require the same capacity as the number of times the \gls{vnf} is visited by that procedure. Multiple \glspl{vnf} can be deployed in a physical node and the total used computational capacity of all the \glspl{vnf} deployed in that physical node cannot exceed its maximum capacity units (i.e. $30$ units). Physical nodes are connected as a mesh topology. The links' propagation delay between physical nodes is set to $5 ms$ for all links. Whereas the processing time each \gls{vnf} takes to process each request of the procedure is randomly assigned between $0.5 ms$ to $1 ms$ based on uniform distribution. The parameters used in the model along with their corresponding values are summarized in Table~\ref{tab:ModelParameters}.
    \begin{table}[h!]
        \centering
        \caption{Parameters used in the Model}
        \label{tab:ModelParameters}
        \resizebox{\columnwidth}{!}{%
        \begin{tabular}{|l||l|}
        \hline
        \textbf{Parameter} & \textbf{Value} \\ \hline \hline 
        Number of physical nodes & $3$ \\ \hline
        Maximum capacity of nodes & $30$ capacity units \\ \hline
        Network connectivity & Mesh topology \\ \hline
        Physical link delay & $5 ms$ \\ \hline
        Physical link maximum bandwidth & $40$ bandwidth units\\ \hline
        Number of \glspl{vnf} & $14$ \\ \hline
        Maximum capacity of \gls{vnf} instance & $10$ capacity units \\ \hline
        \glspl{vnf} base capacity & $1$ capacity unit \\ \hline
        Maximum \gls{vnf} traffic allowed & $2$ (variable in some experiments) \\ \hline
        \glspl{vnf} delay unit & Random between $1000$ and $2000$ packets/sec \\ \hline
        Number of instances per \gls{vnf} & $4$ \\ \hline
        Number of Procedures & $4$ \\ \hline
        Allowed delay for procedure & $1$ second \\ \hline
        Number of slices & $2$ \\ \hline
        \end{tabular}%
        }
    \end{table}
            
\begin{figure}[h!]
        \centering
        \includegraphics[width=8cm, height=6cm] {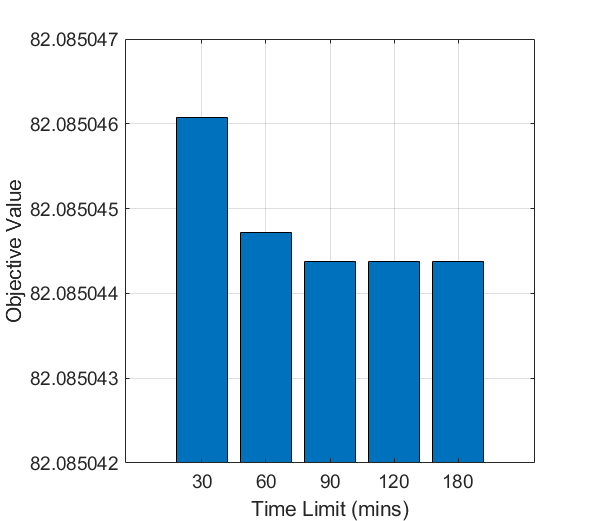}
        \caption{Objective value as a function of different time limits}
        \label{fig:Obj_Val}
    \end{figure}

    \textbf{The simulation time limit:} The \gls{scip} solver is used with two parameters, the maximum number of threads used by the solver and the time limit to solve the model. In these experiments, the maximum number of threads is set to $6$ threads in order to run multiple experiments at the same time. However, we noticed that \gls{scip} only used a single thread at any given time while switching between these threads during the run (i.e. \gls{scip} did not use all $6$ threads concurrently). The time limit, on the other hand, is set in order to obtain a sub-optimal solution from the model in a timely manner. Limiting the time is also considered by previous studies \cite{Coelho2021} to avoid the long time that the model could take to solve a problem with a high number of input parameters. Figure \ref{fig:Obj_Val} compares the objective value obtained as a function of  multiple values of the time limit. When the time limit is set to $30$ minutes, the model provided the highest objective value. As the time limit increases, the objective value starts to saturate. So in light of these results, the time limit is set to $3$ hours for all subsequent experiments and this same value is already used in the literature \cite{Coelho2021}. 

    \textbf{The implemented scenario:} A simple network scenario is implemented in order to obtain and analyze the results from the proposed optimization model. A total of two slices are considered and each slice consists of three procedures. Slice one requires registration with \gls{amf} re-allocation, handover, and authentication procedures. However, slice two requires general registration, handover, and authentication procedures. These procedures are described in the next section. The number of  procedures sourced externally and the maximum \gls{vnf} traffic are varied across the conducted experiments. Table \ref{tab:Imp_Scenario} summarizes the network configuration of the studied scenario.
    \begin{table}[h!]
        \centering
        \caption{Implemented Scenario}
        \label{tab:Imp_Scenario}
        \resizebox{\columnwidth}{!}{%
        \begin{tabular}{|l||l|}
        \hline
        Number of slices & Two \\ \hline \hline
        Procedures for Slice\# $1$ & 1) Registration with \gls{amf} re-allocation procedure \\ & 2) Handover procedure \\ & 3) Authentication procedure \\ \hline
        Procedures for Slice\# $2$ & 1) General registration procedure\\ & 2) Handover procedure\\ & 3) Authentication procedure \\ \hline
        Number of external procedures & Variable \\ \hline
        Maximum \gls{vnf} traffic capacity & Variable \\ \hline
        \end{tabular}%
        }
    \end{table}
    
\section{The Implemented 5G Procedures} 
\label{sec:5g_proc}
    Although our model can support all existing \gls{5g} procedures defined by \gls{3gpp} in \cite{3GPP2021}, only four procedures are implemented in this work to test the visibility and correctness of the model. These procedures are the general registration procedure, registration with \gls{amf} re-allocation procedure, handover procedure, and authentication procedure. In fact, implementing more procedures would enlarge the time needed to get results out of the model. In this section, those implemented procedures are described briefly and the sequence of their serving \glspl{vnf} is provided. More details on these procedures can be found in \gls{3gpp} technical specification 23.502 \cite{3GPP2021}.
    
    
    \textbf{General Registration Procedure:}
    This procedure enables the \gls{ue} to register with the \gls{5g} network to receive services. The \gls{ue} can perform this procedure in different scenarios like the initial registration to join the network, the emergency registration to use the emergency services, etc. 
    The sequence of \glspl{vnf} used by this procedure is as follows: $UE {\rightarrow} \, RAN \,{\rightarrow} \, New\,AMF \, {\rightarrow} \, Old\,AMF \, {\rightarrow} \, New\,AMF \, {\rightarrow} \\ AUSF \, {\rightarrow} \,UDM \, {\rightarrow}  \, New\,AMF \, {\rightarrow} \, UDM \,  {\rightarrow} \, New\,AMF \, {\rightarrow} \\ SDM \, {\rightarrow} \, New\,AMF \, {\rightarrow} \, SDM \, {\rightarrow} \,  New\,AMF {\rightarrow} PCF {\rightarrow} \\ New\,AMF {\rightarrow} \, SMF {\rightarrow} New\,AMF {\rightarrow} UE {\rightarrow} New\,AMF$. Since the \gls{ue} and the \gls{ran} are not actual \glspl{vnf} but do appear in the sequence, we remove the \gls{ue} and the \gls{ran} from the beginning of the procedures' sequence while implementing the procedures in our model. More details on this limitation are explained in section \ref{sec:limit}.
            
    \textbf{Registration with \acrshort{amf} Re-Allocation Procedure:}
    In this procedure, the initial \gls{amf} redirects the registration-related traffic to the target \gls{amf}. For instance, this can happen when the initial \gls{amf} cannot serve the \gls{ue}, so a change in the \gls{amf} is required in this case. 
    One important thing to mention here is that multiple types of \glspl{amf} can be seen in the sequence of \glspl{vnf}, for instance, initial \gls{amf}, target \gls{amf}, etc. In our model, we consider these variants of \glspl{amf} as different  \glspl{vnf} to ensure that these \glspl{vnf} are deployed separately from each other.
    The sequence of \glspl{vnf} used by this procedure is as follows: $RAN {\rightarrow}\, Initial\,AMF \,{\rightarrow} \,UDM\, {\rightarrow} \,Initial\,AMF\, {\rightarrow}\, NSSF \,{\rightarrow} \\ Initial\,AMF {\rightarrow}\,Old\,AMF \,{\rightarrow}\, Initial\,AMF \,{\rightarrow}\, NRF \,{\rightarrow} \\ Initial\,AMF \,{\rightarrow}\, RAN \,{\rightarrow}\,Initial\,AMF \,{\rightarrow}\,Target\,AMF$.
    
    

    \textbf{Handover Procedure:}
    Handover is another important procedure that takes place in cellular networks due to the mobility of the \glspl{ue}. It can also be carried out due to other reasons like load balancing or achieving \gls{qos} requirements. Despite there being other more complex variants of the handover procedures specified by \gls{3gpp} in \cite{3GPP2021}, however, we only consider the simple version of the handover in this work which is called "Xn based inter NG-\gls{ran} handover without \gls{upf} re-allocation". 
    The sequence of \glspl{vnf} used in this procedure is as follows: $Target\,RAN \,{\rightarrow}\, AMF\, {\rightarrow} \,SMF\, {\rightarrow} \,UPF\, {\rightarrow} \,SMF\, {\rightarrow}\, Source\\ RAN \,{\rightarrow} \, Target\,RAN \,{\rightarrow} \,SMF\, {\rightarrow} \,AMF \,{\rightarrow\,}  Target\,RAN\\ \, {\rightarrow} Source\,RAN$.
     
    \textbf{Authentication Procedure:}
    \gls{3gpp} defines two protocols to be used for authentication procedure, \gls{eap-aka} and  5G-\gls{aka}. In this study, we only implement the \gls{eap-aka}. The selection of the authentication protocol is performed by \gls{udm}/\gls{arpf} depending on the \gls{supi} of the \gls{ue} \cite{Project2010}. The authentication procedure can be performed as part of other procedures such as the registration procedure or  UE-triggered service request procedure.
    The sequence of \glspl{vnf} serving this procedure is as follows: $ ARPF\,{\rightarrow} \,UDM \,{\rightarrow} \,AUSF\, {\rightarrow} \,SEAF {\rightarrow} \,UE\, {\rightarrow} \,SEAF\, {\rightarrow} \\AUSF\, {\rightarrow} \, SEAF \, {\rightarrow}\, UE$.

   
\section{Results and Discussion}\label{sec:results}
	This section presents the results obtained from the proposed model. Various performance metrics are considered to test the correctness and  effectiveness of the model including the impact of security constraints, the used capacities of the physical nodes, and the number of activated  \gls{vnf} instances. Additionally, the delay to complete the procedures based on the mapping of \glspl{vnf} to the physical nodes is computed. The main goal of this analysis is to convey the benefits of the proposed security-aware optimization model along with the cost or overhead of prioritizing security. These results only apply to the parameters used and the environment tested. Hence, these results are not a general trend but show how the model could be used.
   
\subsection{Impact of the Exposure Constraint}
    The impact of the exposure security constraint is analysed in terms of the  security goals achieved and the additional overhead that occurred to the network operator. In this set of experiments, the \gls{vnf}’s maximum traffic constraint is disabled. Also, the number of external procedures is varied from $0$ (i.e. no procedure sourced externally) to $4$ (i.e. all procedures sourced externally) in steps of $1$. We assume that  any procedure originated by a \gls{ue} or the \gls{ran} could be defined as a trusted or untrusted procedure. In this work, we consider a particular procedure is externally sourced if it is initiated by an untrusted UE or \gls{ran}. The \gls{hplmn} (i.e., the operator) makes the final decision of whether a particular procedure is identified as trusted or untrusted based on, for example, the identities of the access network and/or visited network. Additionally, the home operator may consider a set of \glspl{ue} or visited networks not sufficiently secure, however, the home operator policy may depend on reasons not related to security features of the connecting \gls{ue} or \gls{ran} to categorize them into trusted or untrusted.
    
\begin{figure}[!h]
        \centering
        \includegraphics[width=9cm, height=8cm] {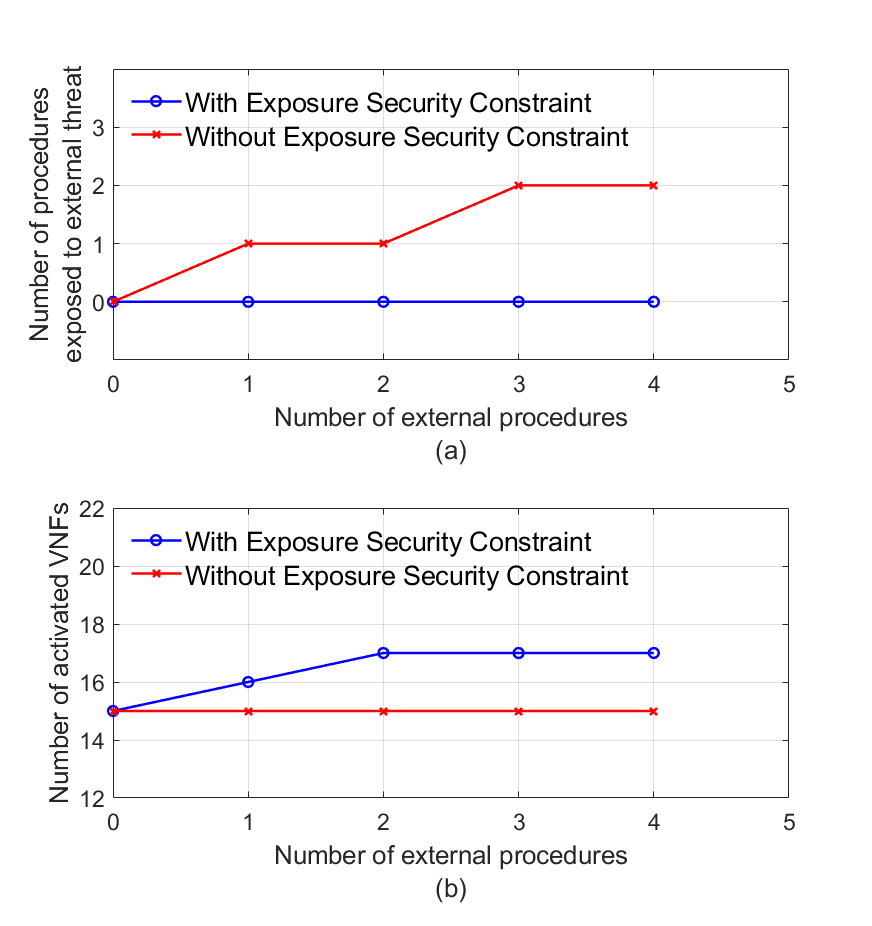}
        \caption{Impact of exposure constraint}
        \label{fig:Expo_Const}
    \end{figure}

    Figure~\ref{fig:Expo_Const}(a) compares the effect of increasing the number of external procedures on the number of procedures exposed to external threats. Here, we check whether the first \gls{vnf} of an external procedure is shared with other procedures. If it is shared, then the other procedures sharing the same \gls{vnf} would be exposed to external threats as well. The assumption here is that the first \gls{vnf} of the external procedure can be a target of attacks from a malicious \gls{ue} or a rouge \gls{gnb} or other sources. As a consequence, the other procedures and slices served by that \gls{vnf} would be exposed to the same external threats. As mentioned earlier in this paper, the exposure security constraint ensures that the first \gls{vnf} of an external procedure is not shared with any other procedures and slices.
    As one can see from Fig.~\ref{fig:Expo_Const}(a), with exposure constraint enabled, as the number of external procedures is increased, the number of procedures exposed to the threat remains at zero. However, with the exposure constraint disabled, more procedures sourced externally result in more procedures exposed to external threats as represented by the red curve. As seen in this figure, there is no increase in the number of procedures exposed to threats when the number of external procedures increases from $1$ to $2$ and from $3$ to $4$ as well. The reason behind this is that the first \gls{vnf} of a procedure considered external is not being used by another procedure, which does not impact the results. Hence the results presented here depend greatly on the implemented scenario.

    Figure~\ref{fig:Expo_Const}(b) presents the cost of including security in the network in terms of the number of activated \gls{vnf} instances. When there is no external procedure, the number of activated \glspl{vnf} is the same with and without the security constraint. As the number of external procedures increases, the result without exposure constraint stays constant. However, with the exposure security constraint turned on, the number of activated \glspl{vnf} would increase. This is because the constraint will ensure that the first \gls{vnf} of the external procedure is not shared with any other procedure resulting in more \gls{vnf} instances activated. More \gls{vnf} activated will directly impact the required capacity which will increase the cost for the network operator. Similar to the previous result, with the exposure constraint, there is no increase in the number of activated \glspl{vnf} when the procedures sourced externally are increased from $2$ to $4$. This can be attributed to the fact that the model already activated separate \glspl{vnf} for different procedures, hence, no new \glspl{vnf} need to be initiated. Another reason could be that the first \gls{vnf} of the external procedure is not used by another procedure which makes no change to the result.
        
\subsection{Impact of the Maximum \gls{vnf} Traffic Constraint}
\begin{figure}[!h]
        \centering
        \includegraphics[width=9cm, height=8cm] {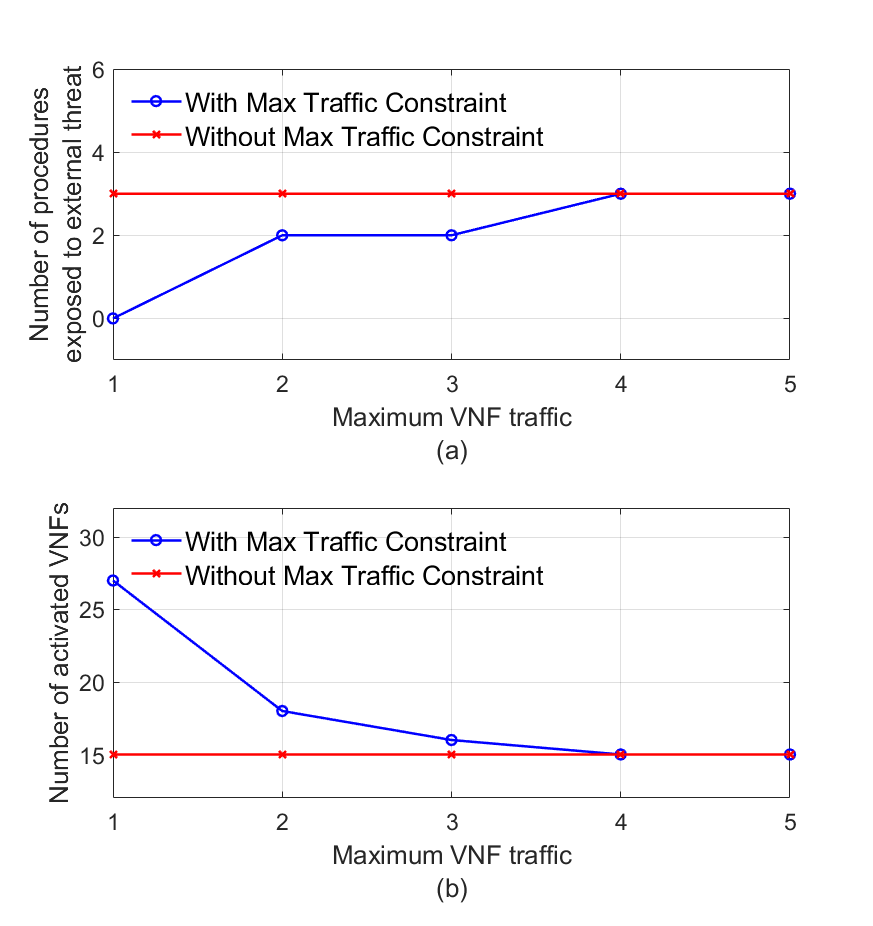}
        \caption{Impact of maximum \gls{vnf} traffic constraint}
        \label{fig:Max_VNF_Const}
    \end{figure}

    To evaluate the impact of the maximum \gls{vnf} traffic constraint, the exposure security constraint is disabled and only one procedure is assumed to be sourced externally. Figure \ref{fig:Max_VNF_Const}(a) shows the benefit of using this constraint. In this set of experiments, the maximum allowed \gls{vnf} traffic is ranging from $1$ to $5$ in steps of one. Also, each \gls{vnf} is set to require one capacity unit to serve one procedure. The maximum \gls{vnf} traffic simply means the number of procedures that the \gls{vnf} can serve. Since one procedure is assumed as external, we consider a procedure exposed to the threat if it shares any \gls{vnf} with the external procedure.
    Based on this assumption, without the maximum traffic constraint, the number of procedures exposed to external threats is constant at $3$ as shown in Fig. \ref{fig:Max_VNF_Const}(a).  With the \gls{vnf}  maximum traffic constraint enabled, the number of exposed procedures is zero initially. This is attributed to that each procedure is mapped to a unique \gls{vnf} and there is no sharing. However, as the maximum limit of \gls{vnf} traffic increases, the number of exposed procedures also increases until it becomes similar to the results without the maximum traffic constraint as shown in Fig.~\ref{fig:Max_VNF_Const}(a).

    Figure~\ref{fig:Max_VNF_Const}(b) shows the cost of implementing this security constraint. The figure shows that when the maximum \gls{vnf} traffic increases, the number of activated \gls{vnf} instances stays constant at $15$. However, with the maximum traffic constraint enabled, the number of initiated \gls{vnf} instances is $27$ when the maximum \gls{vnf} traffic is set to $1$. This in turn requires more capacity and resources by the network operator. The total activated \glspl{vnf} is reduced when the maximum \gls{vnf} traffic limit is increased until it merges with the result of the other scenario (i.e. without the maximum traffic security constraint).
            
\subsection{Physical Node Capacity}
\begin{figure}[!h]
        \centering
        \includegraphics[width=9cm, height=7cm] {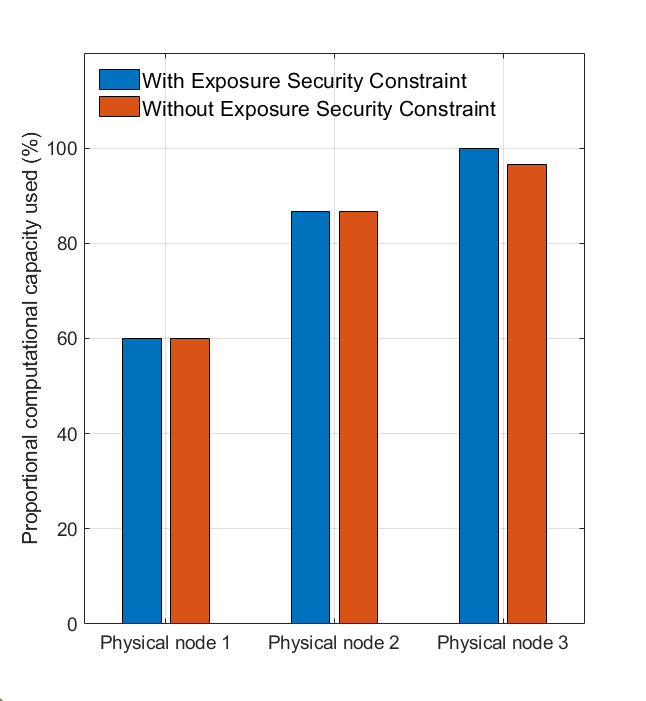}
        \caption{Physical node capacity used}
        \label{fig:Phy_cap}
    \end{figure}

\begin{figure}[!h]
        \centering
        \includegraphics[width=10cm, height=8cm] {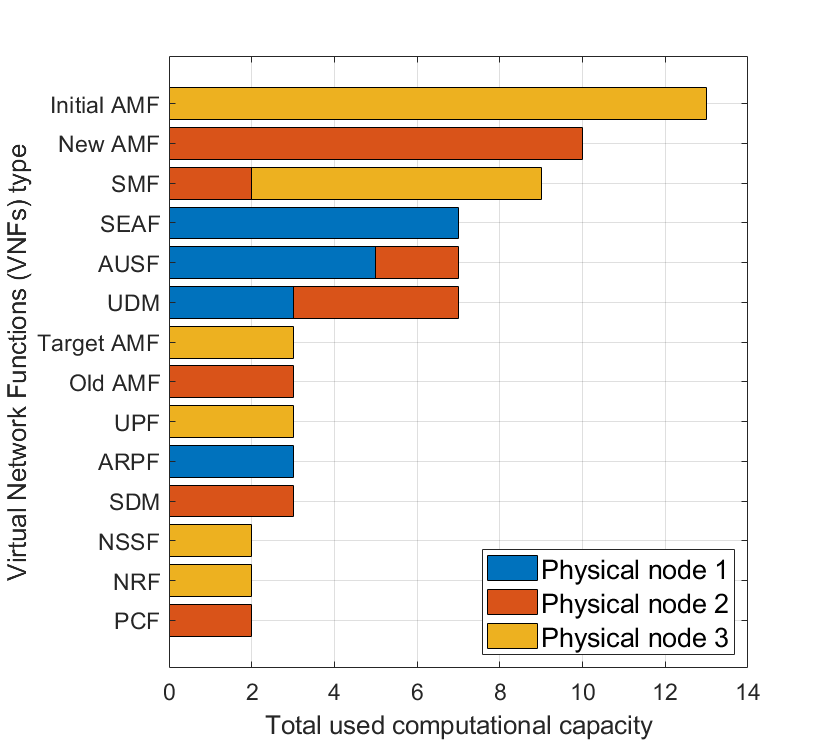}
        \caption{Physical node capacity used by each \gls{vnf} type}
        \label{fig:Phy_cap_VNF}
    \end{figure}

    This subsection shows the amount of physical node capacity required to activate the \glspl{vnf} that meet the slices requirements. For this set of experiments, the maximum \gls{vnf} traffic limit is set to $2$, and only the registration with \gls{amf} re-allocation procedure is set as an external procedure. Figure \ref{fig:Phy_cap} shows the proportional computational capacity used for each physical node. In this experiment, the maximum \gls{vnf} traffic constraint is enabled and the results are obtained with and without the exposure constraint. As shown in the figure, physical nodes $1$ and $2$ consume the same amount of capacity  either with or without the security constraint. However, the capacity of physical node $3$ consumed is $100\%$   with the exposure constraint and about $97\%$ without the security constraint. The major takeaway from this figure is that the extra overhead of the security constraints is not huge if the network operators select moderate security requirements. 
    To scrutinize this further, the total computational capacity of the physical nodes used by each \gls{vnf} is reported in  Fig.~\ref{fig:Phy_cap_VNF}. It can be seen from the figure that the top three \glspl{vnf} that use most of the capacity are the initial \gls{amf}, new \gls{amf}, and \gls{smf}. The initial \gls{amf} and new \gls{amf} are only deployed in physical nodes $3$ and $2$, respectively. The \gls{smf} is mainly initialized in physical node $3$ but another instance of the \gls{smf} is also deployed in physical node $2$. This observation also gives an indication of the \glspl{vnf} which are mostly used by  \gls{5g} procedures and hence making them critical to be protected from threats.

    
\subsection{VNF Instance Capacity}
\begin{figure}[!h]
        \centering
        \includegraphics[width=9cm, height=5cm] {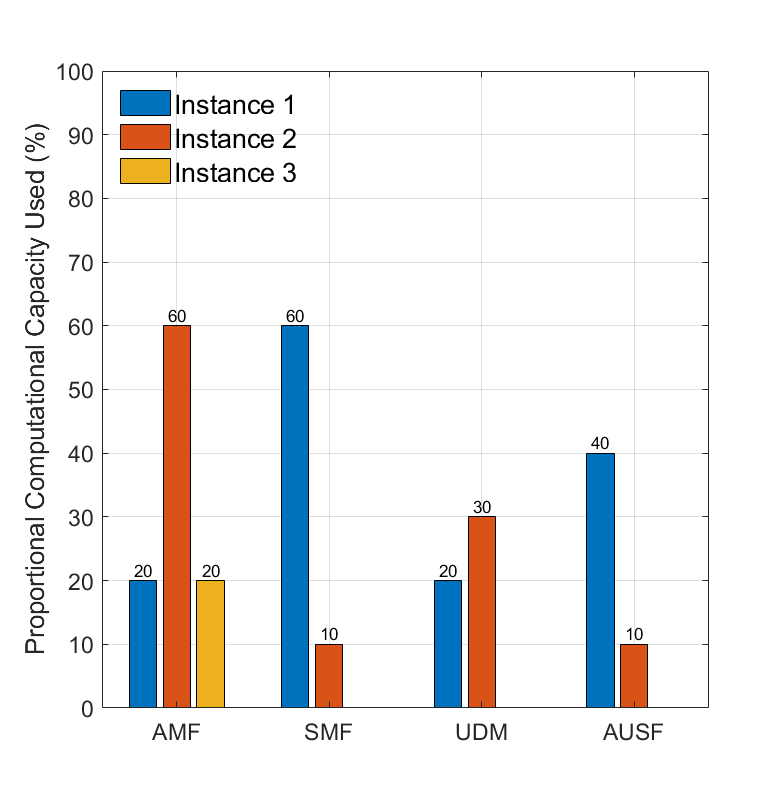}
        \caption{Proportional computational  capacity used by each VNF instance excluding the base capacity}
        \label{fig:vnf_inst_cap}
    \end{figure}

    In this subsection, we show the capacity used by \gls{vnf} instances and their utilization. Figure \ref{fig:vnf_inst_cap} shows the proportional computational capacity used (excluding the base capacity) by each \gls{vnf} instance out of its predefined maximum capacity. Here we only present the results of \glspl{vnf} with more than one instance activated. Both security constraints are enabled in this experiment. It can be seen from the figure that three instances of the \gls{amf} are initiated. The  \gls{amf} consumes the highest total capacity of the initialized instances among all \glspl{vnf}. The \gls{smf}, \gls{udm}, and \gls{ausf} come next with two instances activated for each. 
    
\begin{figure}[!h]
        \centering
        \includegraphics[width=7.5cm, height=5cm] {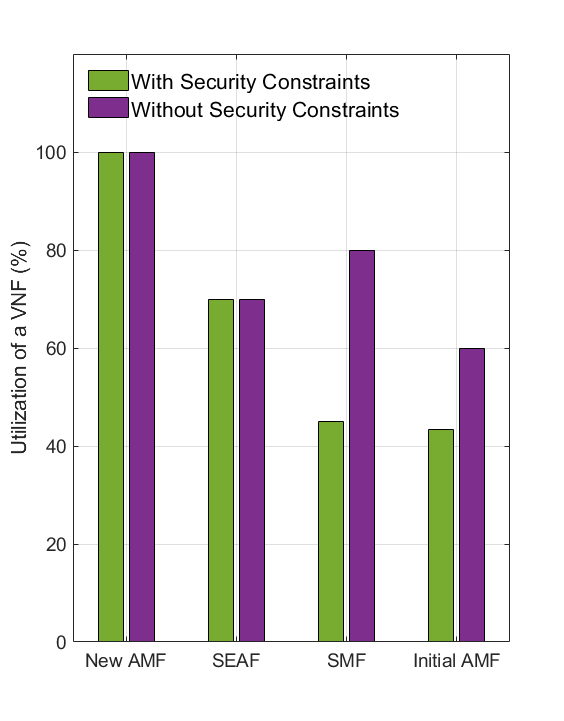}
        \caption{Utilization of \glspl{vnf}}
        \label{fig:vnf_util}
    \end{figure}

    Figure \ref{fig:vnf_util} shows the utilization of \glspl{vnf} with and without the security constraints. The utilization is computed by taking the average of the proportional capacity used across all instances of a particular \gls{vnf} type.
     As shown in Fig.~\ref{fig:vnf_util} the utilization of the new \gls{amf}, for example, is at $100\%$. One important point to mention here is that the utilization of the \gls{smf} and initial \gls{amf} when the security constraints are enabled is lower than when they are disabled. The reason behind this is that the limit on the maximum \gls{vnf} traffic and the exposure constraint results in more \gls{vnf} instances activated, reducing the overall \gls{vnf} utilization. However, this is a trade-off to make between protecting the network against threats and achieving higher utilization.
    			
\subsection{Procedure Delay}
\begin{figure}[!h]
        \centering
        \includegraphics[width=9cm, height=7cm] {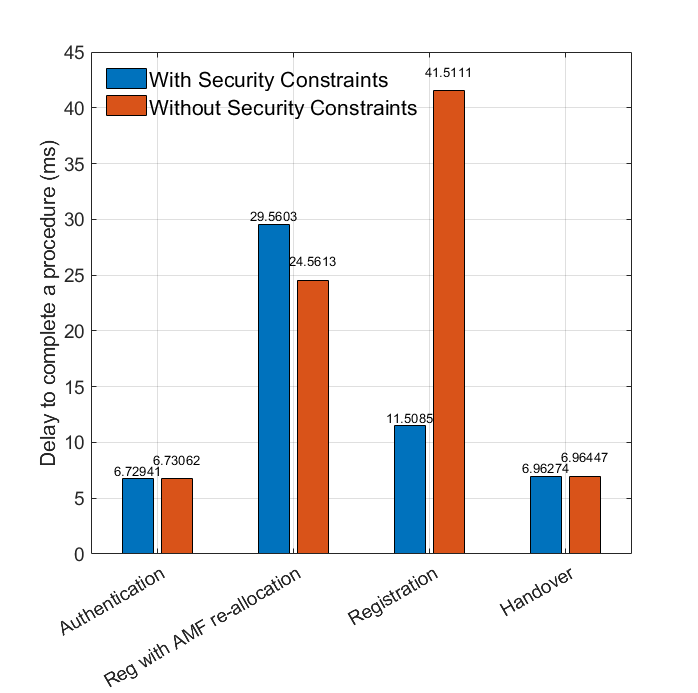}
        \caption{Delay to complete a procedure}
        \label{fig:delay}
    \end{figure}

    Lastly, we calculate the time it takes for each procedure to be completed. A delay average is reported if the same procedure is used across slices. The experiments are performed with both security constraints enabled and when they are disabled. As Fig.~\ref{fig:delay} shows, the delay to complete the authentication and handover procedures is around $7 ms$ and it is almost the same for both scenarios (i.e. with and without the security constraints). The registration with the \gls{amf} re-allocation procedure takes $5 ms$ more when the security constraints are enabled. However, the major difference in the delay is in the registration procedure. The delay to complete the procedure in the experiment without the security constraints is $30 ms$ more than the delay in the experiment when the security constraints are enabled. This difference in the delay is because, for the results without the security constraints, some \glspl{vnf} were arbitrarily deployed in different physical nodes resulting in extra propagation delay that contributes to the total delay. 
    Another reason could be the limited run time for the model, which only provides a sub-optimal solution once the time limit is reached. Additionally, there is a trade-off between sharing  \glspl{vnf}  and the delay as a lower number of  \glspl{vnf}  does not always guarantee less delay due to multiple factors that can influence the delay.

\section{Limitations}\label{sec:limit}
    This section attempts to identify some limitations of the proposed model in this work and the way that it is planned to deal with them. These limitations are summarized in the following points:
    \begin{itemize}
        \item \textbf{\gls{ue} and \gls{ran} assignments:} The first limitation of the current implementation is considering the \gls{ue} and the \gls{ran} as  \glspl{vnf}. Since \gls{ue} and \gls{ran} are also part of \gls{5g} procedures explained in section \ref{sec:5g_proc}, the sequence of network entities that are involved during the procedure also includes the \gls{ue} and the \gls{ran}. Currently, we do not distinguish between \gls{ue} or \gls{ran} and other 5G  \glspl{vnf}  in the implementation.
        In order to get around this issue, we remove the \gls{ue} and the \gls{ran} from the beginning of the procedures' sequence of  \glspl{vnf}. This is done to ensure that the exposure constraint does not consider the \gls{ue} or the \gls{ran} as the first \gls{vnf} of procedures set to be sourced externally. Therefore, removing the \gls{ran} or \gls{ue} from the beginning of the sequence will guarantee that a 5G \gls{vnf} will be the first \gls{vnf} of this procedure. Another technique we employed is to assign zero base and processing capacities to the  \gls{ue} and the  \gls{ran}. As a result, the model will map the \gls{ue} and  \gls{ran} to the physical nodes (similar to other \glspl{vnf}) but their capacities will not be impacting the total capacity of the physical node. This is done to ensure the \gls{ue} and the  \gls{ran} contribute to the procedure delay without consuming the computational capacities of physical nodes.

        \item \textbf{Model time limit:} Another limitation of the current model is that we limit the run time of the model to $3$ hours. This is done to obtain the results from the model in a timely manner. Once the time limit is reached, the model will provide the best solution obtained so far in terms of the objective function. Since there is a limit on the model run time, the results presented in this study might not be optimal.
    \end{itemize}
           
\section{Conclusion} \label{sec:conclusion}
    In this work, we propose an optimization-based security-aware \gls{vnf} sharing model for \gls{5g} systems. The goal of the proposed model is not only to enable the efficient mapping of the \glspl{vnf} to maximize their utilization but also to isolate slices by not sharing their critical \glspl{vnf} to enhance security. For this, we introduce a systematic way to decide whether to share  a particular \gls{vnf} or not. To do so, two security constraints were defined in the proposed model; \gls{vnf}’s maximum traffic and \gls{vnf} exposure constraints. The overall goal of the objective function is to minimize the computational capacity required and the total procedure delay. The numerical results of the model are obtained using the four standard \gls{5g} procedures with actual \glspl{vnf}. The results show the advantage of using the security constraints in terms of securing the network slices, procedures, and \glspl{vnf} by limiting the sharing of critical \glspl{vnf}. The use of security constraints introduces additional costs to the network operators in the form of more capacity used. However, the use of security constraints will ensure the protection of critical network infrastructure from external threats such as, for example,  \gls{ddos} attacks. 

\section{Acknowledgement}
    This work was supported by the Natural Sciences and Engineering Research Council of Canada (NSERC) and TELUS Communications through the Collaborative Research and Development (CRD).


\bibliographystyle{IEEEtran} 
\bibliography{library}
\end{document}